\documentclass[%
 aip,
 amsmath,amssymb,
 reprint,%
]{revtex4-1}

\usepackage{dcolumn}
\usepackage{bm}
\usepackage[utf8]{inputenc}
\usepackage[T1]{fontenc}
\usepackage{mathptmx}
\usepackage{etoolbox}
\usepackage[dvipdfmx]{xcolor}
\usepackage[normalem]{ulem}
\usepackage[pdftex]{graphicx}

\makeatletter
\def\@email#1#2{%
 \endgroup
 \patchcmd{\titleblock@produce}
  {\frontmatter@RRAPformat}
  {\frontmatter@RRAPformat{\produce@RRAP{*#1\href{mailto:#2}{#2}}}\frontmatter@RRAPformat}
  {}{}
}
\makeatother
\begin{document}

\title{Development of a bipolar 50~V output digital-to-analog converter system for ion-shuttling operations}

\author{T. Oshio}
\affiliation{Graduate School of Engineering Science, Osaka University, 1-3 Machikaneyama, Toyonaka, Osaka, Japan.}

\author{R. Nishimoto}
\affiliation{Graduate School of Engineering Science, Osaka University, 1-3 Machikaneyama, Toyonaka, Osaka, Japan.}
\affiliation{National Institute of Information and Communications Technology, 588-2, Iwaoka, Nishi-ku, Kobe, Hyogo, Japan.}

\author{T. Higuchi}
\affiliation{Institute for Integrated Radiation and Nuclear Science, Kyoto University, 2, Asashiro-Nishi, Kumatori-cho, Sennan-gun, Osaka, Japan.}

\author{K. Hayasaka}
\affiliation{National Institute of Information and Communications Technology, 588-2, Iwaoka, Nishi-ku, Kobe, Hyogo, Japan.}
\affiliation{Graduate School of Engineering Science, Osaka University, 1-3 Machikaneyama, Toyonaka, Osaka, Japan.}

\author{K. Koike}
\affiliation{e-trees. Japan, Inc., Daiwaunyu Building 2F, 2-9-2 Owadamachi, Hachioji, Tokyo, Japan.}

\author{S. Morisaka}
\affiliation{Center for Quantum Information and Quantum Biology, Osaka University, 1-2 Machikaneyama, Toyonaka, Osaka, Japan.}
\affiliation{QuEL, Inc., Daiwaunyu Building 3F, 2-9-2 Owadamachi, Hachioji, Tokyo, Japan.}

\author{T. Miyoshi}
\affiliation{QuEL, Inc., Daiwaunyu Building 3F, 2-9-2 Owadamachi, Hachioji, Tokyo, Japan.}
\affiliation{Center for Quantum Information and Quantum Biology, Osaka University, 1-2 Machikaneyama, Toyonaka, Osaka, Japan.}
\affiliation{e-trees. Japan, Inc., Daiwaunyu Building 2F, 2-9-2 Owadamachi, Hachioji, Tokyo, Japan.}

\author{R. Ohira}
\affiliation{QuEL, Inc., Daiwaunyu Building 3F, 2-9-2 Owadamachi, Hachioji, Tokyo, Japan.}

\author{U. Tanaka}
\altaffiliation[Author to whom correspondence should be addressed: ]{utako@ee.es.osaka-u.ac.jp}
\affiliation{Graduate School of Engineering Science, Osaka University, 1-3 Machikaneyama, Toyonaka, Osaka, Japan.}
\affiliation{Center for Quantum Information and Quantum Biology, Osaka University, 1-2 Machikaneyama, Toyonaka, Osaka, Japan.}
\affiliation{National Institute of Information and Communications Technology, 588-2, Iwaoka, Nishi-ku, Kobe, Hyogo, Japan.}

\date{\today}        

\begin{abstract}

The quantum charge-coupled device (QCCD) is one of the notable architectures to achieve large-scale trapped-ion quantum computers.
To realize QCCD architecture, ions must be transported quickly while minimizing motional excitation. 
High-voltage sources are necessary to achieve such high-quality ion transport through a high secular frequency.
In this study, we report the development of a field programmable gate array (FPGA)-based digital-to-analog converter (DAC) system with an output voltage range of $\pm 50$~V and demonstrate its effectiveness in ion transport operations.
The device provides 16-channel analog output, maximum update rate of 16 mega updates per second (MUPS), slew rate of 20 V/µs, and bandwidth of > 200 kHz.
By optimizing the voltage sets with quadratic programming, we experimentally confirmed that this DAC system can achieve more than twice the secular frequency attainable when its output range is restricted to $\pm10$~V, which is consistent with the fact that scaling all electrode voltages by a factor of 5 will scale the secular frequency by $\sqrt{5}$.
Since the output range of many commercially available DACs is commonly limited to $\pm10$~V, this increase is effective for ion shuttling operations, such as transport, split and merge.
The developed DAC system has potential to increase the speed of ion transport thereby reducing processing times in QCCD-based quantum computers.
\end{abstract}

\maketitle

\section{\label{sec:intro}Introduction}

Trapped-ion qubits~\cite{bruzewicz2019trapped} are one of the most promising physical platforms for realizing quantum computers due to their long coherence times~\cite{wang2021single} and high-fidelity quantum gate operations~\cite{PhysRevLett.117.060504, PhysRevLett.117.060505, loschnauer2024scalable}.
So far, quantum computers with tens of trapped-ion qubits have been developed~\cite{chen2024benchmarking, PRXQuantum.2.020343,decross2024,PhysRevX.13.041052}. 
However, scaling to larger systems remains a significant challenge. 

One promising approach for realizing large-scale trapped-ion quantum computers is the quantum charge-coupled device (QCCD) architecture~\cite{Wineland1998,Ekielpinski2002}.
In this architecture, time-dependent voltages are applied to the trap electrodes to shuttle trapped-ion qubits between multiple operational regions, such as those for quantum gates, memory, state preparation, and measurement.
Since the initial proposal by Wineland \textit{et al.}~\cite{Wineland1998}, numerous advancements have been made, and the QCCD architecture continues to be actively studied in the community~\cite{Kaushal2020, pino2021demonstration, PhysRevX.13.041052, PRXQuantum.4.040313}.

\begin{figure*}
\includegraphics[width=0.95\textwidth]{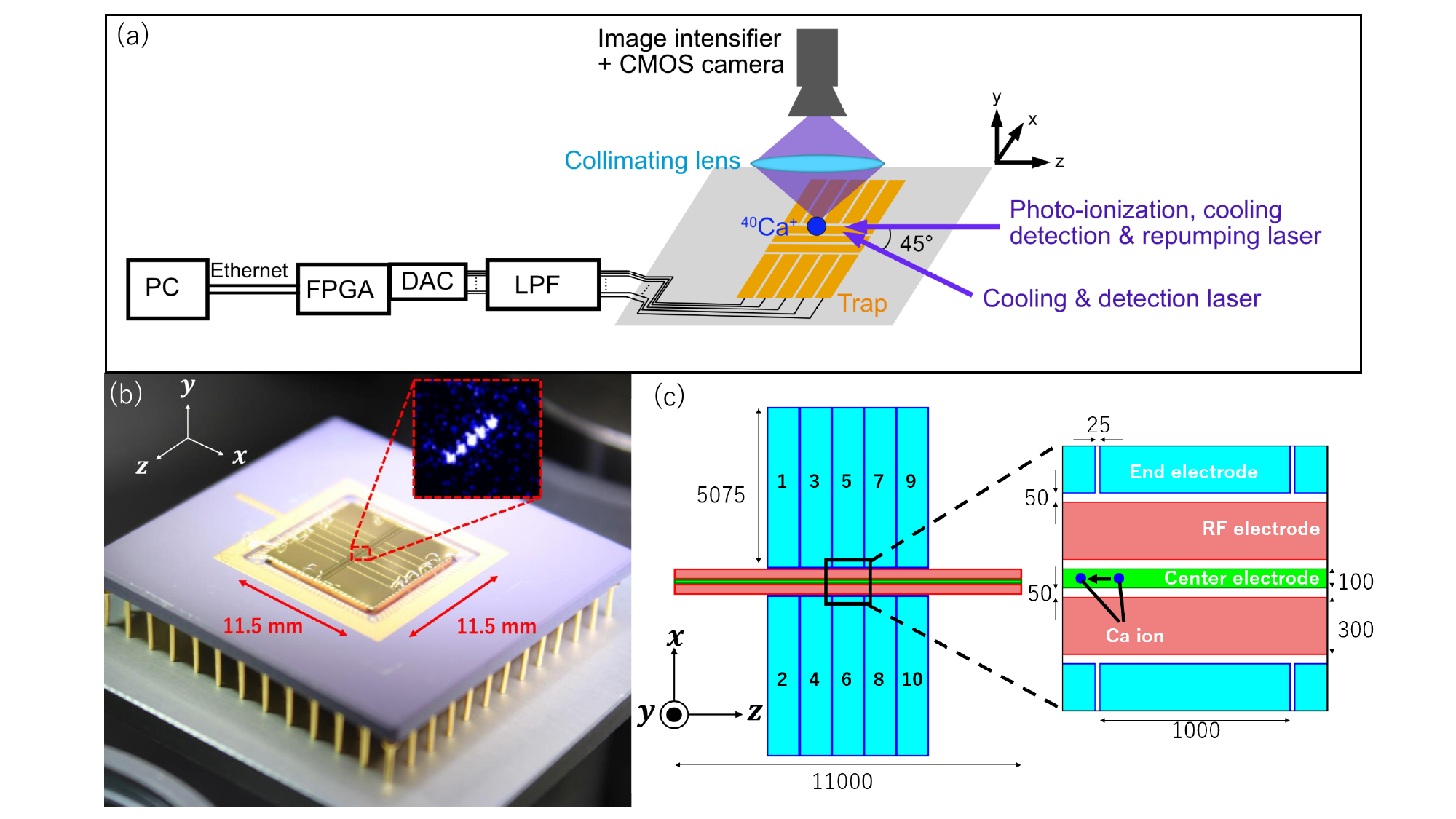}
\caption{\label{fig:PlanerIonTrap} 
Overview of the experimental setup. 
(a) Block diagram of the experimental setup.
The Low-Pass Filter (LPF) is a first-order Butterworth filter with a cutoff frequency of 2~kHz.
The cooling and detection laser refers to the 397~nm laser, and the laser from the same ECDL is split into two beams using a beam splitter.
The photo-ionization laser refers to the 375~nm laser and 423~nm laser.
The repumping laser refers to 866~nm laser.
(b) Planar ion trap used in this study with a fluorescence image of five ${}^{40}\mathrm{Ca}^+$ ions. 
The size of the trap chip is 11.5~mm square and placed in vacuum system.
(c) Schematic of the trap structure.
The black arrow between the blue circles represents the path of ion transport.
The units of the scale are in $\mu$m.}
\end{figure*}

A key technical challenge in realizing a QCCD-based quantum computer is achieving fast ion transport.
The processing time of a QCCD trapped-ion quantum computer is largely dominated by ion cooling, the longest operation in the sequence.
This is followed by shuttling operations, which highlights the importance of increasing the speed of the shuttling ions without heating of the motional state of the ions~\cite{pino2021demonstration}.

Typically, the speed of the shuttling ions is constrained by the secular frequency~\cite{Reichle2006,Nizamani_2011}, when the ion is transported adiabatically.
Therefore, it is important to increase the secular frequency to transport the ion faster.
Since the secular frequency is determined by the trap structure and the applied voltage, improving the output range of digital-to-analog converter (DAC) is effective.
In addition, split/merge operations require to generate a double potential, and to achieve it, a high frequency or high voltages is usefull\cite{home2006,Palmero2015,Ruster2014,Kaufmann2014}.
To date, DAC system with an output range of $\pm 40$~V has been developed and demonstrated for shuttling 
operations~\cite{Kaushal2020}.

In this study, we report the development of a field programmable gate array (FPGA)-based DAC system, which is capable of generating an output range of $\pm 50$~V.
We integrated this DAC system into our trapped-ion setup and conducted experimental assessments to evaluate its performance.
Using the DAC system, we trapped a $^{40}\rm{Ca}^+$ ion with voltage sets optimized with quadratic programming\cite{Frank1956} in a planar trap and measured a secular frequency exceeding $2\pi\times500$~kHz.
In contrast, when the DAC system's output range was limited to $\pm 10$~V, which is a typical output limitation of commercially available DACs, the achievable secular frequency was restricted to $2\pi\times240$~kHz.
The results are consistent with the fact that scaling all electrode voltages by a factor of 5 will scale the secular frequency by $\sqrt{5}$.
The increase in secular frequency achieved using this DAC system helps expand the adiabatic region for ion transport.
This advancement represents a step towards faster and more precise ion transport in large-scale QCCD quantum computers.

This paper is organized as follows.
Section~\ref{sec:method} describes the experimental equipment and protocol for ion-shuttling operations.
Section~\ref{sec:DevelopmentofnewDACsystem} provides detailed information on the developed DAC system and its performance.
In Section~\ref{sec:Result}, we experimentally assess the performance of the developed DAC system.
Finally, Section~\ref{sec:Conclusion} presents the conclusions of this study.

\section{\label{sec:method}Experimental method}

In this section, we first provide an overview of the experimental setup, focusing on the trap electrode and laser configuration.
Next, we explain the method of optimizing the voltage sets for the ion-shuttling operations and that of measuring the secular frequency of the trapped ions.

\subsection{\label{sec:planariontrap}Planar ion trap}

The overview of the experimental setup is shown in Fig.~\ref{fig:PlanerIonTrap}(a).
All experiments presented in the study were performed using the planar ion trap shown in Fig.~\ref{fig:PlanerIonTrap}(b). 
This trap is housed in an ultra-high vacuum system with a vacuum level of approximately $6.7\times10^{-8}$~Pa.
The planar trap consists of two RF electrodes and eleven DC electrodes, including ten end electrodes and a center electrode, as shown in Fig.~\ref{fig:PlanerIonTrap}(c). 
The dimensions of each electrode are as follows: the end electrodes are 5075~$\mu$m$\times$1000~$\mu$m, the RF electrodes are 300~$\mu$m$\times$11000~$\mu$m, and the center electrode is 100~$\mu$m$\times$11000~$\mu$m.
The gap between the RF and DC electrodes is $50~\mu$m, and the gap between adjacent DC electrodes is $25~\mu$m.
The RF and end electrodes provide confinement along the $x$-direction, while the RF, center and parts of the end electrodes contribute to confinement along the $y$-direction. 
The end electrodes provide confinement along the $z$-direction. 
The ions are trapped above the center electrode.

We applied RF voltage at $2\pi\times24.31$~MHz to the RF electrodes, generated by an oscillator and amplified to approximately 200~$\mathrm{V_{pp}}$ using an amplifier and a helical resonator.
As shown in Fig.~\ref{fig:PlanerIonTrap}(a), all of the DC voltages were controlled by the DAC system (see section~\ref{sec:DevelopmentofnewDACsystem} for details of the system).
The signals from the DAC system passed through a low-pass filter (LPF) to suppress noise and smooth rapid voltage changes.
When initially trapping ions in the experiment, the typical trapping voltages for the end electrodes were [$V_{1}$,$V_{2}$,$\cdots$,$V_{10}$]= [$0$,$0$,$10$,$10$,$0$,$0$,$10$,$10$,$0$,$0$]~V.
Here, $V_{1}$,$V_{2}$,$\cdots$,$V_{10}$ refer to the voltages applied to the end electrodes labeled in Fig.~\ref{fig:PlanerIonTrap}(c).
The typical trapping voltage for the center electrode was 3~V.
Using these trapping conditions and an analytical solution\cite{House2008} of the trap potential, which was produced by rectangular electrode, we estimated the secular frequencies to be approximately $(\omega_x,\omega_y,\omega_z)\approx2\pi\times(2.0, 2.9, 0.34)$~MHz, where $\omega_i$ denotes the secular frequency along the $i$-axis, and the trapping height from the trap surface was approximately 190~$\mu$m.

\subsection{\label{sec:laser}Laser beam configuration and fluorescence detection of ions}

In Fig.~\ref{fig:PlanerIonTrap}(a), the laser beam configuration and detection system are depicted.
In this study, all experiments were performed using ${}^{40}\mathrm{Ca}^+$ ions.
${}^{40}\mathrm{Ca}^+$ ions are isotope-selectively produced using a photoionization method~\cite{gulde2001simple, PhysRevA.69.012711}.
Laser beams at 423~nm and 375~nm were applied to a neutral Ca atomic beam.
The ions were Doppler-cooled by applying optical beams at 397~nm and 866~nm.
The 397~nm, 423~nm, and 866~nm lasers were homemade External-Cavity Diode Lasers (ECDL), while a commercially available laser source was used for the 375~nm laser (Nichia NDU4116).
The wavelengths of these laser beams were measured using a wavemeter (HighFinesse WS7), calibrated with a He-Ne laser (Thorlabs HRS015B).
The ions were detected by irradiating the 397~nm laser beam and the 866 nm repump beam onto the ${}^{40}\mathrm{Ca}^+$ ions, exciting the ${}^2S_{1/2}\leftrightarrow{}^2P_{1/2}$ transition and repumping from the $D_{3/2}$ level.
The fluorescence from trapped ions collimated by the lens and amplified with an image intensifier (Hamamatsu Photonics C9016-02) was detected by a CMOS camera (Thorlabs DCC3240M).

\subsection{\label{sec:Voltage Optimization}The Voltage optimization for ion-shuttling operations}
For the ion-shuttling operation, we attempted to find the voltage sets that produce the electrostatic potential $\Phi$ while minimizing the difference between $\Phi$ and target potential $\Phi_{\mathrm{target}}$:
\begin{align}
     \mathrm{Minimize~:~}{\| \Phi -\Phi_{\mathrm{target}}\|}^{2}.
\end{align}
Here, $\Phi$ refers to the potential generated by the applied voltage and $\Phi_{\mathrm{target}}$ is the potential chosen to ensure that the position of null point of the electric field and the value of the second derivative are as desired.
For each potential, we consider only the vicinity of trapped ion in the range [$z_{\mathrm{min}},z_{\mathrm{max}}$].
The settings of $z_{\mathrm{min}}$ and $z_{\mathrm{max}}$ are made for each optimization.
The definitions will be provided later.
To solve such a minimization problem and determine the voltage sets, we used quadratic programming.~\cite{Frank1956}.
Quadratic programming is a type of nonlinear optimization method that minimizes an objective function:
\begin{align}
    \label{eq:objective function}
    \mathrm{Minimize~:~} \frac{1}{2}x^\mathsf{T}Px+q^\mathsf{T}x,
\end{align}
where matrix $P$ and vector $q$ are determined by the desired potential (details will be discussed later).
$x$ is a vector, representing the set of voltages, which is defined as
\begin{align}\label{eq:x-matrix}
 x^\mathsf{T} = (V_{a},V_{b},V_{c},V_{d},V_{e}).
\end{align}
In this optimization, we imposed a constraint that each pair of electrodes with the same $z$-component was set to be the same voltage, such that $V_{1}=V_{2}=V_{a}$, $V_{3}=V_{4}=V_{b}$, $V_{5}=V_{6}=V_{c}$, $V_{7}=V_{8}=V_{d}$, $V_{9}=V_{10}=V_{e}$.
We optimized the voltages for ion transport from the start point to the end point by adjusting the voltages $N$ times.
The start position is represented as step~0, and the end point as step~$N$. 
$N$ is any natural number and here we set $N = 10$ to avoid noise with frequency components in the transport waveform close to secular frequency.
At each step~$k$ $(0\leq k \leq N)$, we express the ion position as $z_{k}$, which is corresponding to the null point of the electrostatic potential.
The electrostatic potential $\Phi_{k}$ around $z=z_{k}$ can be approximated by a harmonic potential along the $z$-axis:
\begin{align}
 \label{eq:harmonicpotential}
  \Phi_{k}(z) = \frac{m\omega_{z}^{2}}{2e}(z-z_{k})^{2},
\end{align}
where $m$ is the ion mass and $e$ is the elementary charge.
In this experiment, $z_{k}$ is defined as:
\begin{align}
 \label{eq:shuttlingposition}
  z_{k}= z_0+ (z_{N}-z_{0})\mathrm{sin^{2}}(\frac{\pi k}{2N}),
\end{align}
to smoothen the transport speed changes and thereby, reduce the heating of the motional states of the ion.

To generate the required potential along the $z$-axis, we define the basis function $\phi_{\mathrm{base}_{i}}(z)$ as the potential generated when a unit voltage is applied to each electrode pair $i\in(a,b,c,d,e)$.
The potential at each step is a linear combination of these basis functions, with the optimization range ($L$) for each step~$k$ set as $z_{\mathrm{min}} = z_{k} - L/2$ and $z_{\mathrm{max}} = z_{k} + L/2$.
Here, we set $L=50~\mu$m.
The matrix $v$ and vector $f$ are defined as:
\begin{align}
\label{eq:v-matrix}
v=
\begin{pmatrix} 
  \phi_{\mathrm{base}_{a}}(z_{\mathrm{min}}) & \dots  & \phi_{\mathrm{base}_{a}}(z_{\mathrm{max}}) \\
  \vdots &     & \vdots \\
  \phi_{\mathrm{base}_{e}}(z_{\mathrm{min}}) & \dots & \phi_{\mathrm{base}_{e}}z_{\mathrm{max}})\\
\end{pmatrix}, 
\end{align}
\begin{align}
\label{eq:f-vector}
f = (\Phi_{k}(z_{\mathrm{min}}),\dots,\Phi_{k}(z_{\mathrm{max}})).
\end{align}
Using $v$ and $f$, $P$ and $q$ can be expressed as:
\begin{align}
P=2v\cdot v^\mathsf{T},\\
q=-2v\cdot f^\mathsf{T}.
\end{align}
Additionally, the Full Scale Range (FSR) of the DAC imposes restrictions on $x$, which are incorporated into the optimization by defining $G$ and $h$ as: 
\begin{align}
G=
\begin{pmatrix} 
  1 & 0 & 0 & 0 & 0 \\
  -1 & 0 & 0 & 0 & 0\\
  0 & 1 & 0 & 0 & 0 \\
  0 & -1& 0 & 0 & 0 \\
  0 & 0 & 1 & 0 & 0 \\  
  0 & 0 & -1 & 0 & 0 \\
  0 & 0 & 0 & 1 & 0 \\
  0 & 0 & 0 & -1 & 0 \\
  0 & 0 & 0 & 0 & 1 \\
  0 & 0 & 0 & 0 & -1 \\
\end{pmatrix} 
,h=\frac{\mathrm{(FSR)}}{2}
\begin{pmatrix} 
  1 \\1 \\ 1 \\ 1 \\ 1 \\  1 \\ 1 \\ 1 \\ 1 \\ 1 \\
\end{pmatrix}.
\end{align}
We included these inequality constraints as:
\begin{align}
  \label{eq:inequality constraints}
  Gx\leq h.
\end{align}

\begin{figure*}
\includegraphics[width=0.95\textwidth]{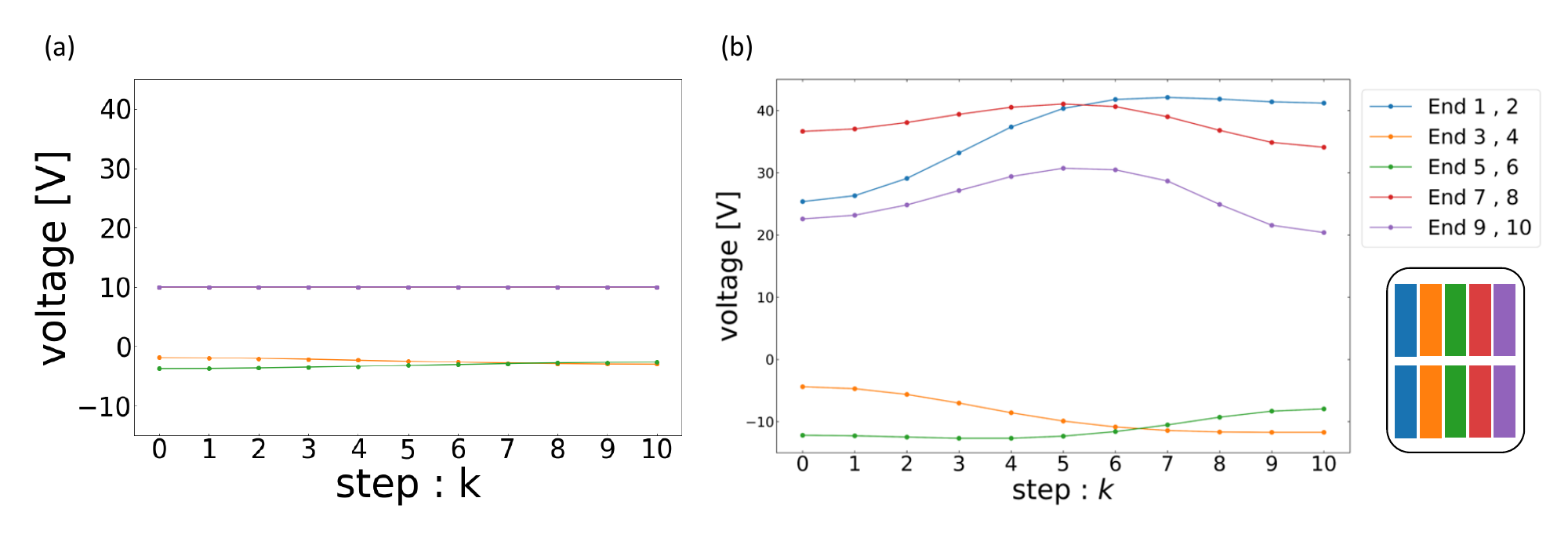}
\caption{\label{fig:optimized voltage}
Optimized voltage sets with (a) an FSR of 20~V (DAC output range of $\pm$10~V) and (b) an FSR of 100~V (DAC output range of $\pm$50~V). 
The schematic in the lower right shows the arrangement of the end electrodes, corresponding to Fig.~\ref{fig:PlanerIonTrap}(c), with colors matching the plot curves and marks.
The voltage sets were calculated using quadratic programming for a constant secular frequency $\omega_z=2\pi\times500$~kHz.}
\end{figure*}
With these parameters, the voltage set $x$ was optimized using common Python package: Python Software for Convex Optimization.
Fig.~\ref{fig:optimized voltage}(a) and Fig.~\ref{fig:optimized voltage}(b) show the optimized result with an FSR of 20~V (DAC output range of $\pm10$~V) and 100~V (DAC output range of $\pm50$~V), respectively.
In both cases, the target secular frequency $\omega_z$ was set as $2\pi\times500$~kHz, with a transport distance of $200$~$\mu$m.
In Fig.~\ref{fig:optimized voltage}(a), the blue, red and purple lines overlapping at approximately +10 V are due to the voltage restriction.
To verify that the optimized voltage sets can operate properly, we performed ion shuttling 10,000 times, with each step duration set to 100~$\mu$s in the presence of cooling laser.

\subsection{\label{sec:measurement}Measurement of secular frequency}
Here, we describe the method we used for measuring the secular frequency of a single trapped ion at each step during the ion-shuttling operation.
The secular frequency was measured by observing the resonance of the ion motion with the imaging system shown in Fig.~\ref{fig:PlanerIonTrap}(a) by sweeping the frequency of an RF excitation and observing an increase in the ion width on resonance.
We applied an oscillating voltage to one electrode of the trap (End 6) to resonantly excite the motion of a single ion\cite{Staanum2008}.
We set the amplitude of oscillating voltage to 30~$\mathrm{mV_{pp}}$.
We swept the frequency of an oscillating voltage over a range of approximately 15~kHz with a step of 0.1~kHz and recorded the ion image to extract the ion width at the same time.
We then fitted the results with a Lorentzian to obtain the secular frequency.

\begin{figure}
\includegraphics[width=0.48\textwidth]{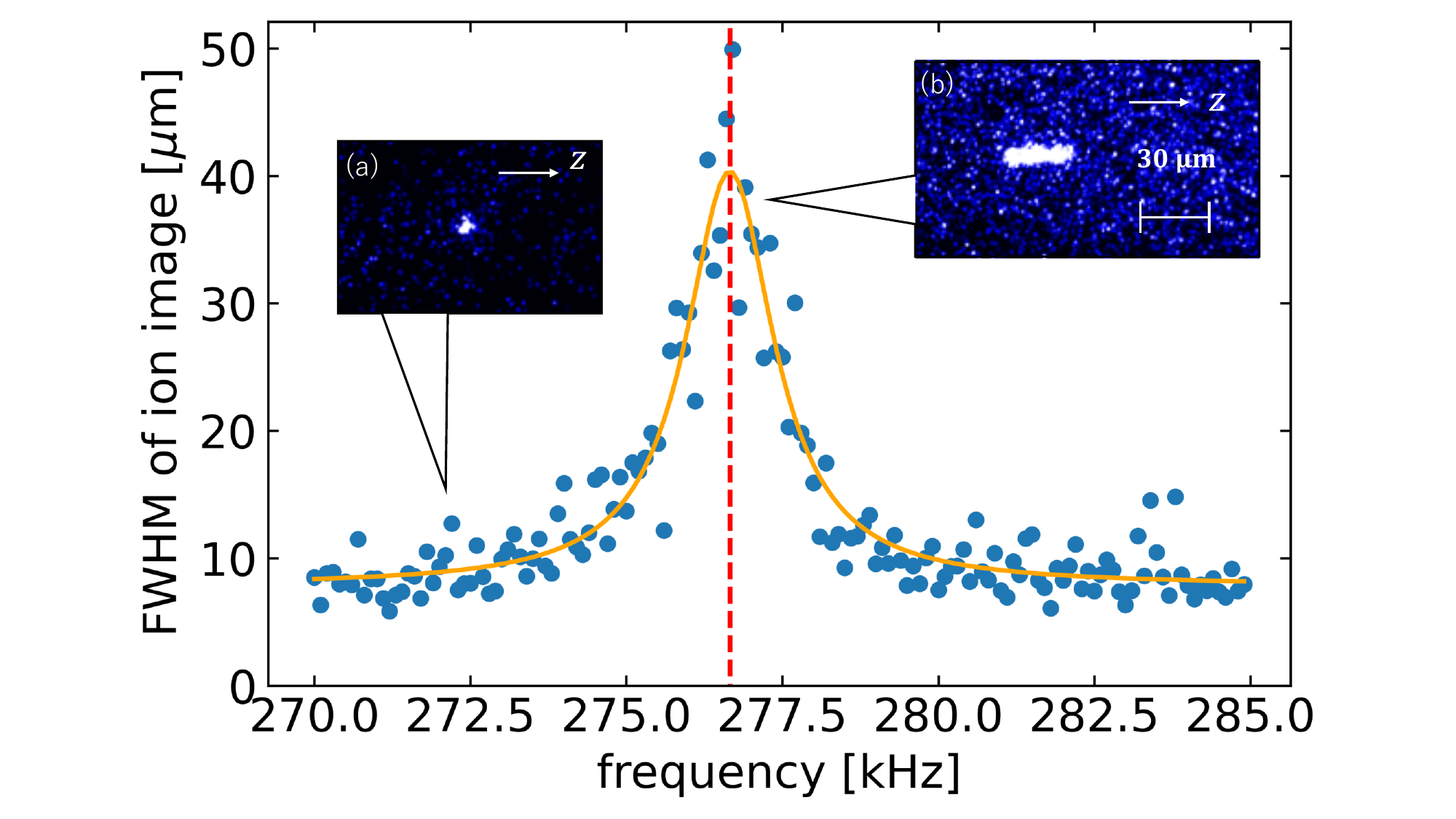}
\caption{\label{fig:sampleofFHWM}
Full Width at Half Maximum (FWHM) of ion fluorescence as a function of the frequency of the signal applied to the end electrode (End 6). The signal frequency was swept in increments of 0.1~kHz. The FWHM data (blue dots) were fitted with a Lorentzian (yellow curve), with the peak position (red dashed line) indicating the secular frequency $\omega_{z}$, determined to be $2\pi\times$276.6~kHz. 
The FWHM of the fitted Lorentzian was 1.72~kHz.
The data corresponds to shuttling step 0 when the voltage output range is limited to $\pm10$~V. The insets show fluorescence images of the ion (a) off-resonance and (b) on-resonance.
}
\end{figure}

The result of measuring the secular frequency is shown in Fig.~\ref{fig:sampleofFHWM}.
The secular frequency was calculated by fitting the ion fluorescence data (blue dots) with a Lorentzian (yellow curve). 
The peak of the fitted curve (red dashed line) was identified as the secular frequency, $\omega_{z}$.
To obtain these results, the experimental sequence was repeated ten times, and the full width at half maximum (FWHM) was determined as the average of the fluorescence profiles.
The ion fluorescence was acquired with the camera exposure time set to 10 ms.
From this measurement, the secular frequency $\omega_{z}$ was determined to be $2\pi\times$276.6~kHz.
\section{Development of a high-voltage DAC system}\label{sec:DevelopmentofnewDACsystem}

\begin{figure}
    \includegraphics[width=0.48\textwidth]{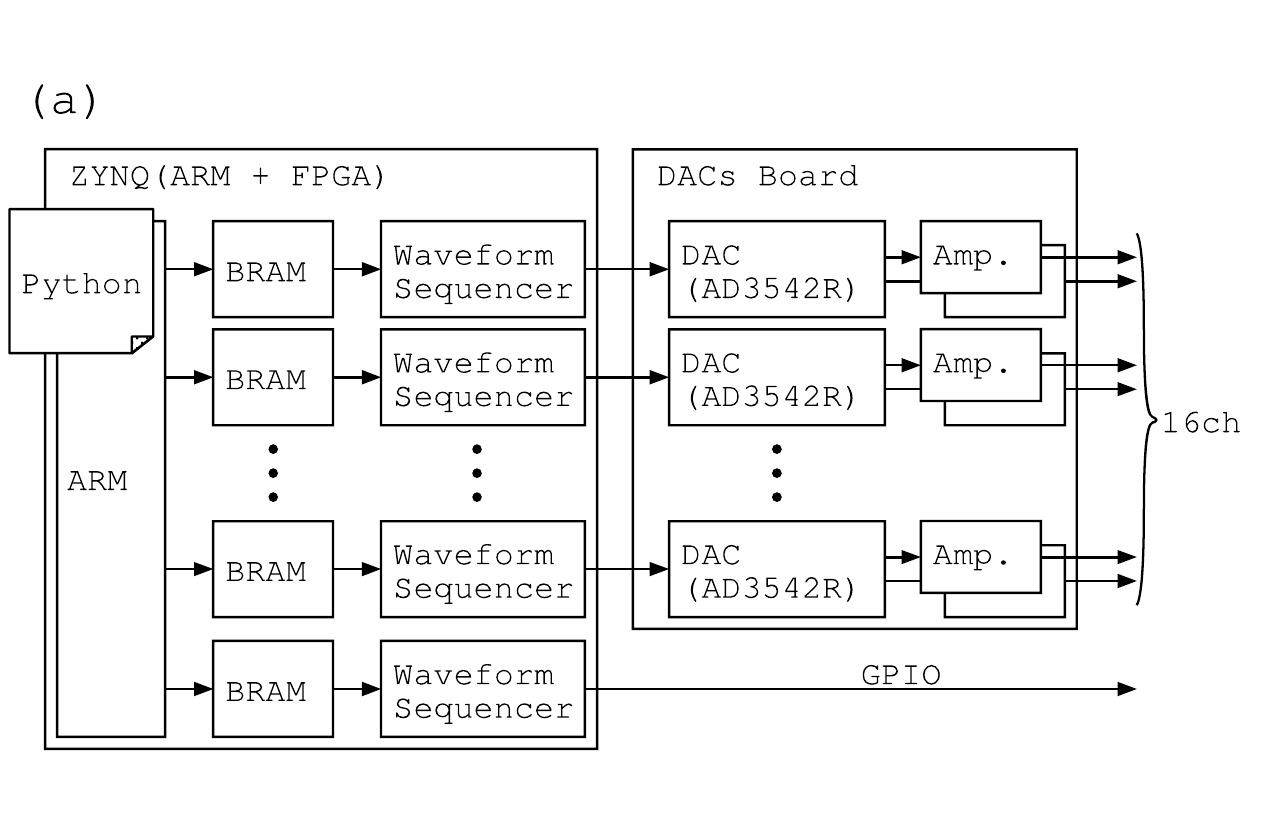}
    \includegraphics[width=0.48\textwidth]{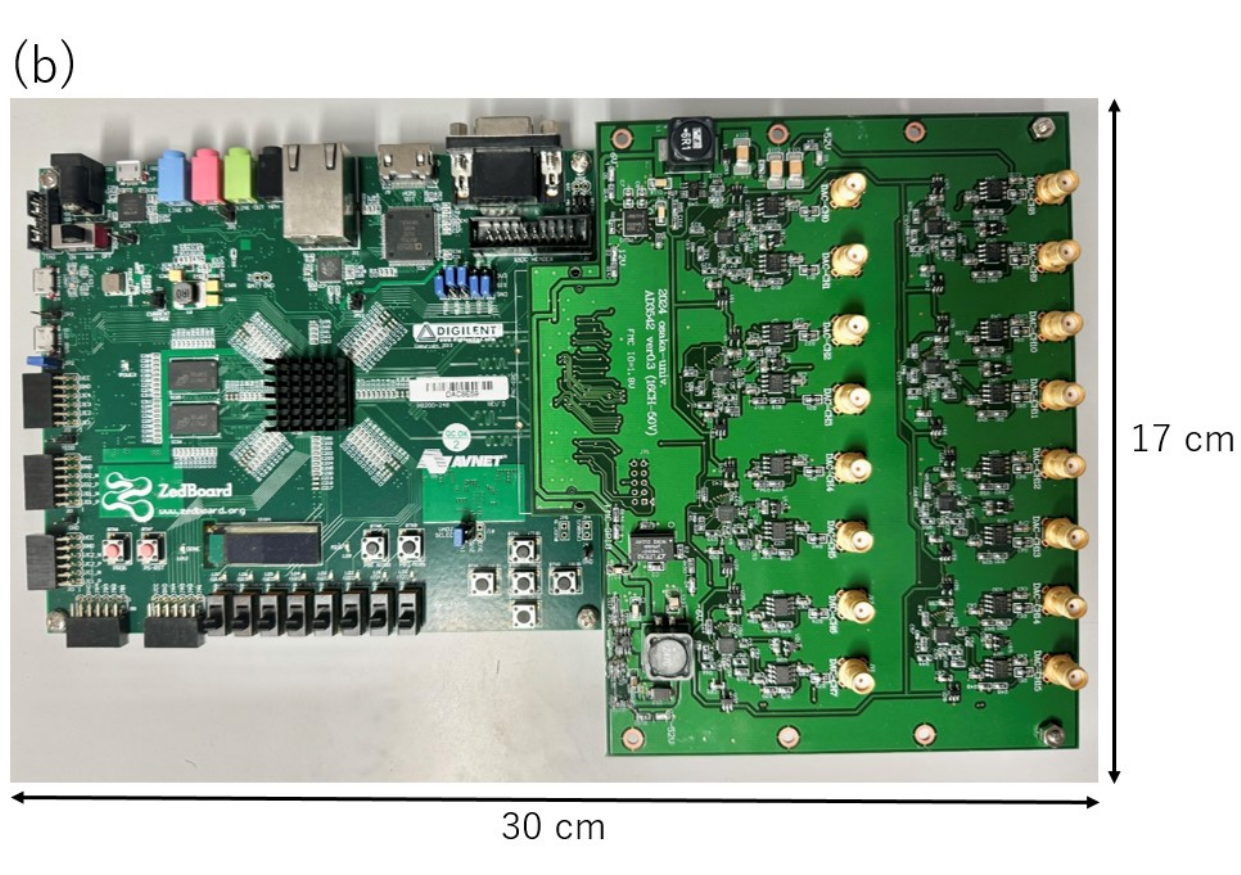}
\caption{\label{fig:controller_overview}
 Overview of the developed DAC system. The system consists of an Zynq SoC board (Zedboard) and a custom-designed analog signal output board. (a) Block diagram of the developed DAC system. (b) Photograph of the Zynq SoC board and an analog output board.}
\end{figure}

The overview of the developed DAC system is shown in Fig.~\ref{fig:controller_overview}.
The system includes a controller and 16 DACs.
The controller is an AMD Zynq 7020 System-on-Chip (Zynq SoC)~\cite{zynq7000}, implemented on a Digilent ZedBoard~\cite{zedboard}. The Zynq SoC consists of an ARM processing system (PS) and programmable logic (PL, also known as an FPGA).
The DACs are AD3542R\cite{ad3542r} models from ADI.
The controller and each DAC communicate via a dual SPI interface.
Output pattern data for the DACs are stored in multiple BlockRAMs (BRAMs) within the PL.
To control the DAC output voltage, the user writes pattern data to the BRAMs using a Python script running on the PS. 
The design specification of the developed DAC system is shown in Table.\ref{tab:analog_spec}.
\begin{table}
    \centering
    \caption{Design specifications of the output signal.}
    \label{tab:analog_spec}
    \begin{tabular}{l c}
    \hline\hline
    Number of channels & 16 \\
    Voltage range      & +/- 50V \\
    Output voltage resolution  & 16-bit or 12-bit \\
    Maximum update rate & 16 MUPS \\
    Maximum threw rate & 20 V/$\mu$s \\
    \hline\hline
    \end{tabular}
\end{table}

\begin{figure}
\includegraphics[width=0.49\textwidth]{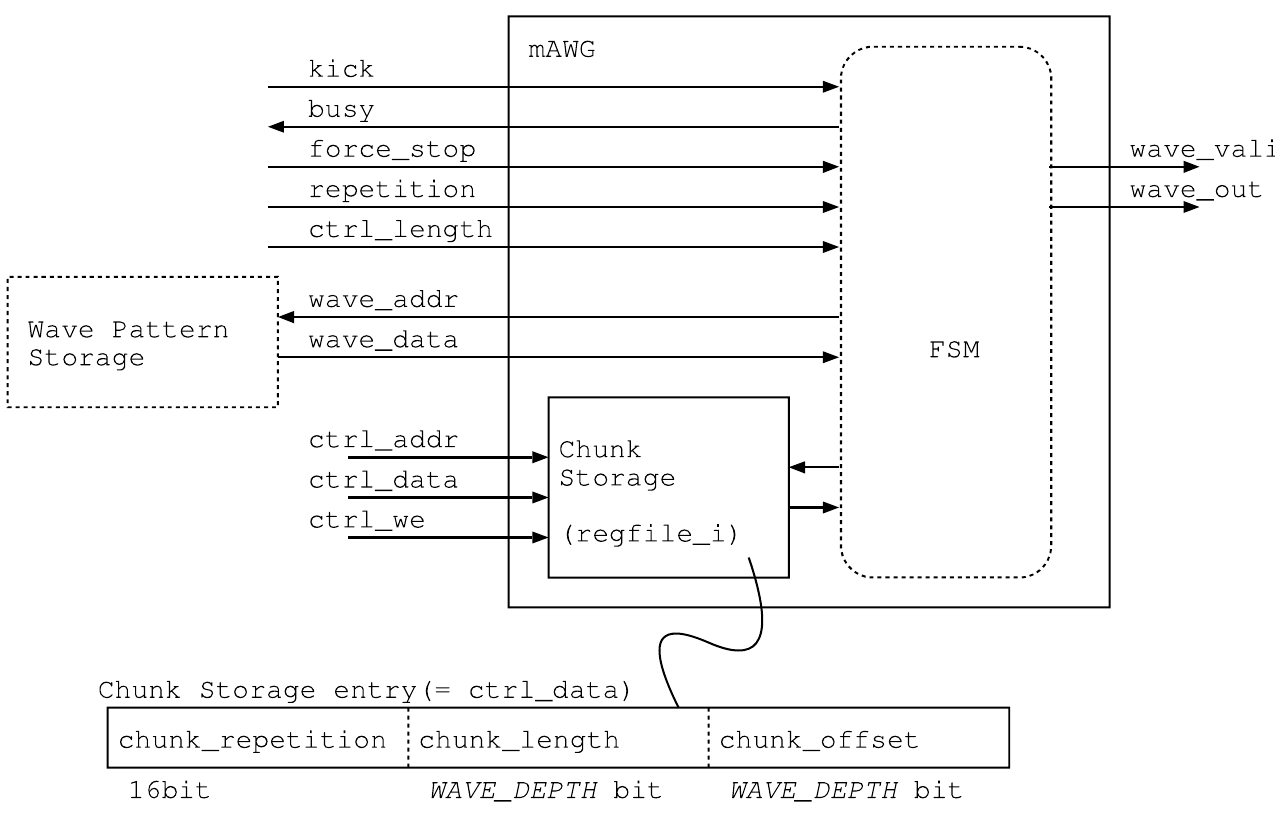}
\caption{\label{fig:sequencer_overview}
Overview of the waveform sequencer. This sequencer generates waveforms by reading data from $\mathtt{Wave~Pattern~Storage}$ according to rules defined in $\mathtt{Chunk~Storage}$. The generated waveform is illustrated in Fig. \ref{fig:wave_definition}.}
\end{figure}

Fig.~\ref{fig:sequencer_overview} shows the internal architecture of the sequencer. 
This sequencer generates waveforms by reading data from the Wave Pattern Storage based on entries in the Chunk Storage. 
The module starts running when the \verb|kick| signal is asserted, during which the \verb|busy| signal remains active. 
The module generates waves according to the specified parameters, but the users can stop the module at any time by asserting the \verb|force_stop| signal. 
Only when the module asserts \verb|wave_valid|, the \verb|wave_out| signal is valid.

\begin{figure}
\includegraphics[width=0.49\textwidth]{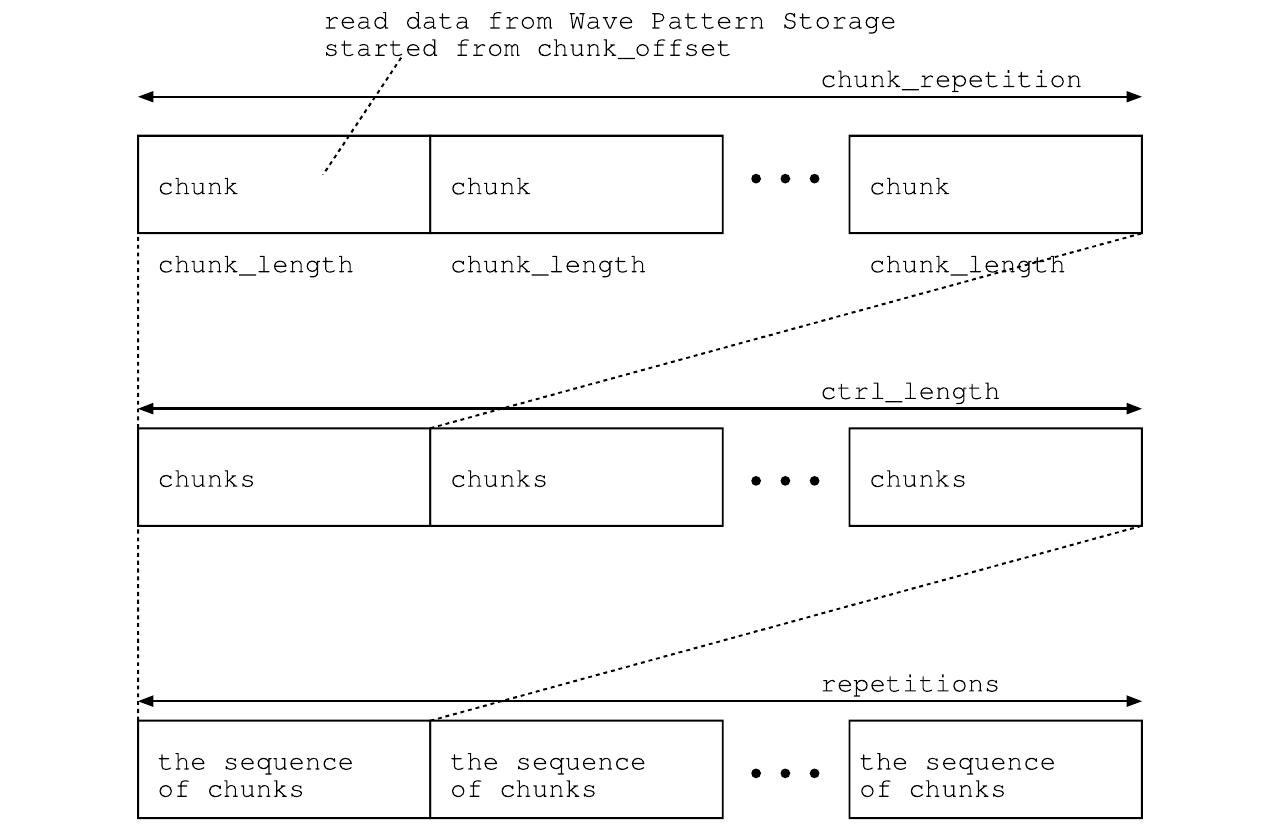}
\caption{\label{fig:wave_definition} 
Overview of the waveform pattern. The memory footprint of the defined waveform is reduced by utilizing repetition patterns within the waveform.}
\end{figure}

\begin{figure}
\includegraphics[width=0.49\textwidth]{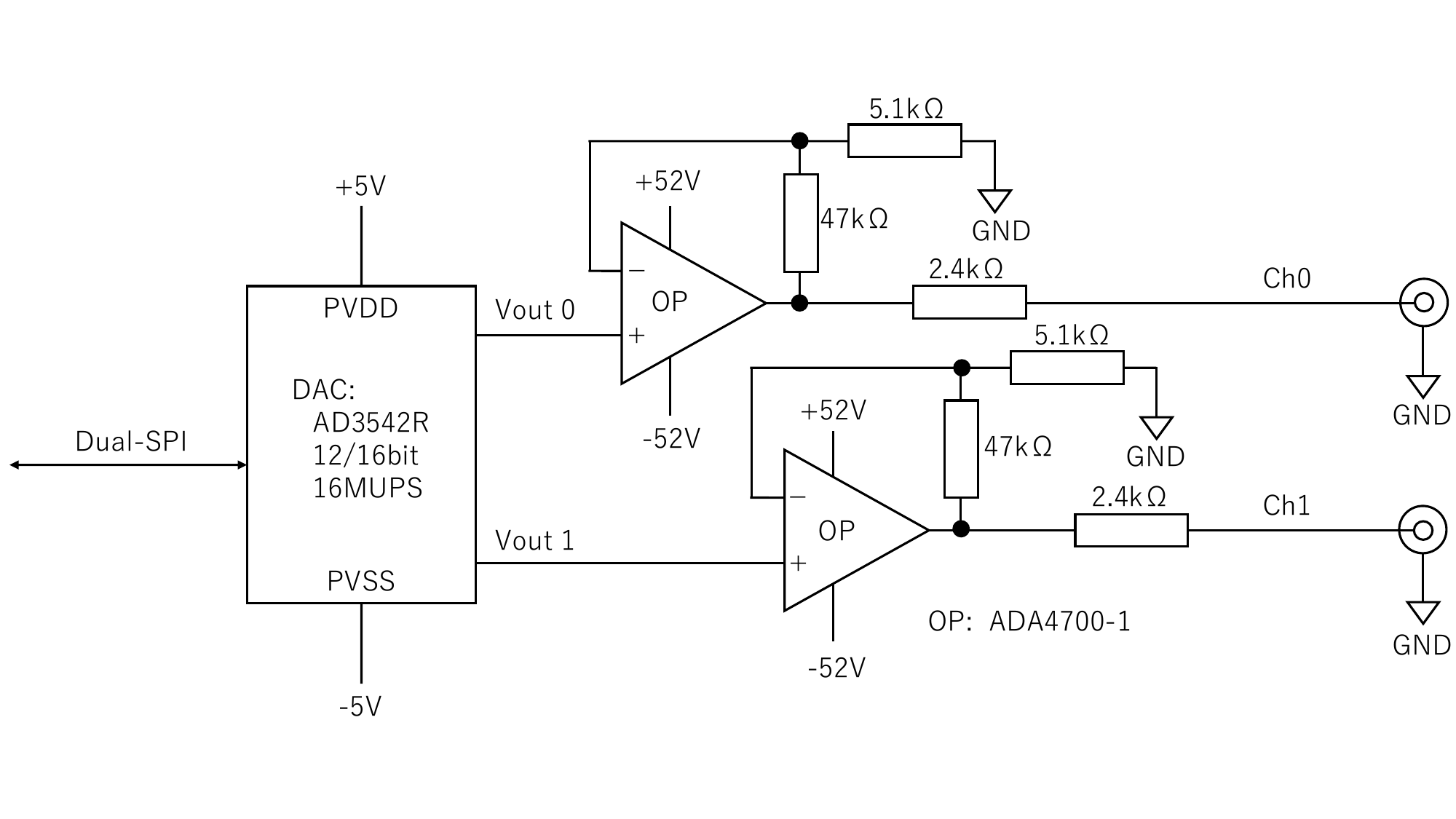}
\caption{\label{fig:ad3542_16ch_circ} 
Schematic of the analog output module used in this study.
The gain of non-inverting amplifier circuit is 10.2.
This non-inverting amplifier circuit contributes to the realization of the DAC output range of $\pm50$~V.}
\end{figure}

The definition of the wave patterns is shown in Fig.~\ref{fig:wave_definition}. 
The waveform is generated by combining multiple chunks defined by the parameters of the Chunk Storage entry. 
Each Chunk Storage entry contains \verb|chunk_length|, \verb|chunk_offset|, and \verb|chunk_repetition|. 
The parameters \verb|chunk_offset| and \verb|chunk_length| specify the start address and length of the Wave Pattern Storage to read for each chunk. 
The parameter \verb|chunk_repetition| specifies the number of repetitions of the chunk. 
The number of entries read from the Chunk Storage is given by \verb|ctrl_length|, and repetition specifies the number of repetitions for reading the sequence of the chunks.

The multiple DACs convert the generated waveform into analog signals, which are then amplified to a range of $\pm$50~V using high-throughput operational amplifiers (OP-amps).
Fig.~\ref{fig:ad3542_16ch_circ} illustrates the schematic of a DAC and two amplifiers. 
The DAC used in this system, the AD3542R, features two output ports, each connected to an independent OP-amps (ADA4700\cite{ada4700}).
While the DAC provides 
slew rate of 100~V/$\mu$s, the overall slew rate of the analog signal output is limited by the OP-amp's slew rate of 20~V/$\mu$s. 

The noise density specification states that the noise density of AD3542R is 30~nV/$\sqrt{\mathrm{Hz}}$ (@10~kHz), while that of ADA4700 is 10~nV/$\sqrt{\mathrm{Hz}}$ (@10~kHz).
Considering that the gain is 10, AD3542R becomes the dominant factor, and the noise density of this system is approximately 300~nV/$\sqrt{\mathrm{Hz}}$ (@10~kHz).
The Total Harmonic Distortion (THD) specifications at 10~kHz are -90~dB for AD3542R and -94~dB for ADA4700, resulting in an overall system THD of approximately -88~dB at 10~kHz.

\section{Result and Discussion}\label{sec:Result}

Using the voltage sets derived in subsection~\ref{sec:Voltage Optimization}, we trapped ${}^{40}\mathrm{Ca}^+$ ions at each step i.e., at each position during ion transport, and measured $\omega_z$ by the ion resonance experiments described in subsection~\ref{sec:measurement}.  

\begin{figure}
\includegraphics[width=0.49\textwidth]{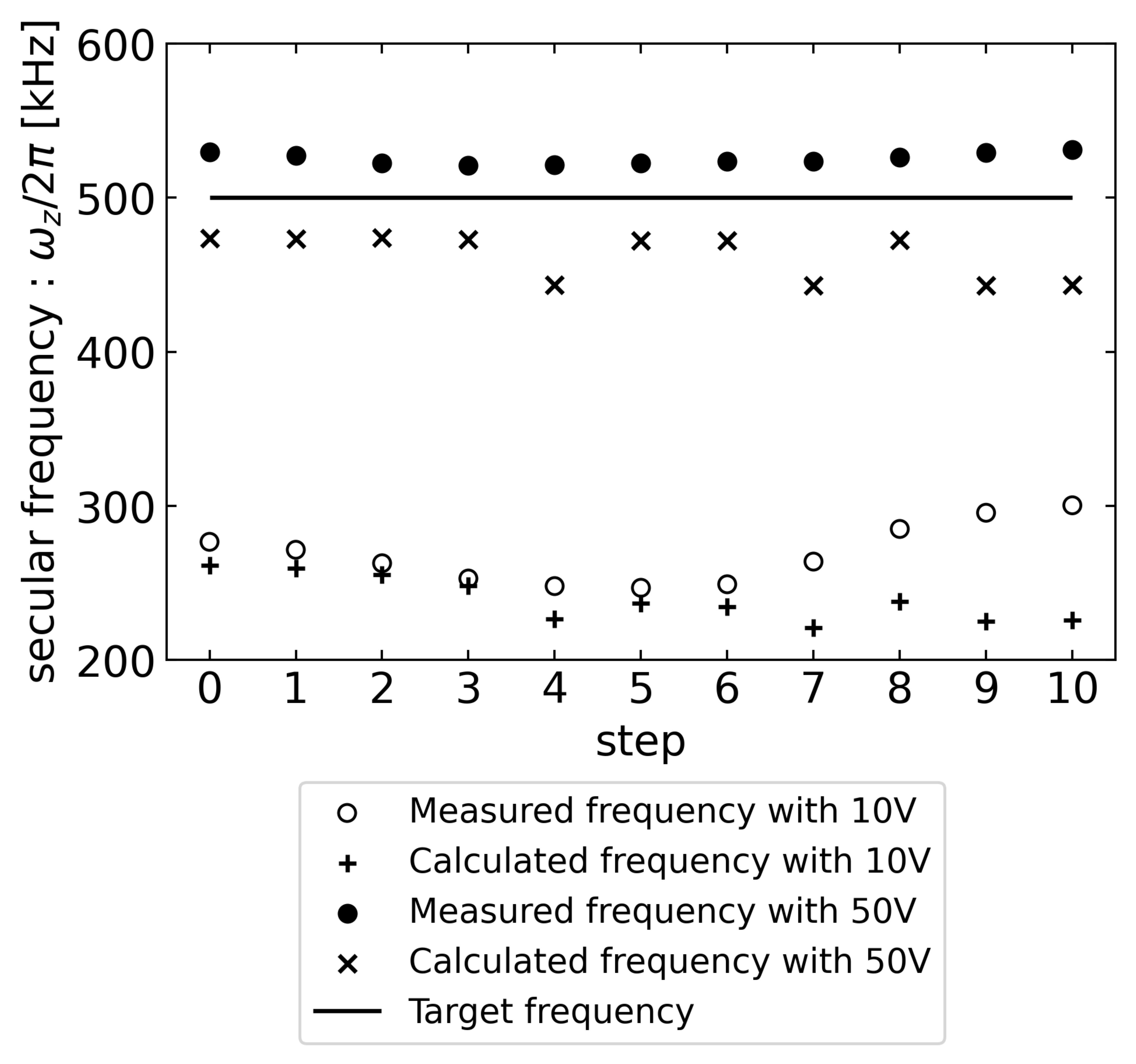}\caption{\label{fig:Secular frequency}
Measured secular frequencies of ions trapped at each step of the shuttling operation using the voltage sets shown in Fig.~\ref{fig:optimized voltage}.
The results are presented for the DAC's output range limited to $\pm10$~V (white dots) and $\pm50$~V (black dots).
The size of the error bars is within the marker size. 
Calculated secular frequencies based on the used voltage sets are also shown for the DAC's output range limited to $\pm10$~V (plus markers) and $\pm50$~V (cross markers). 
The horizontal line indicates the target $\omega_{z}$ used for optimization.}
\end{figure}

The measurement results are shown in Fig.~\ref{fig:Secular frequency}.
The horizontal line at $\omega_{z} = 2\pi\times500$~kHz indicates the optimization target.
It is experimentally observed that the secular frequency cannot achieve the target secular frequency when the DAC output range is limited to $\pm10$~V.
This limitation arises because many of the optimized voltage values shown in Fig.~\ref{fig:optimized voltage}(a) reach the $\pm10$~V limit.
In contrast, with a $\pm50$~V DAC output, the measured secular frequencies closely match the target frequency.
The frequency differences of \textasciitilde 50 kHz between the calculated frequencies and the measured frequencies are suspected to be caused by potential imperfections, such as those caused by the contact potential induced by Ca atom deposition on the electrodes.

\begin{figure}
\includegraphics[width=0.49\textwidth]{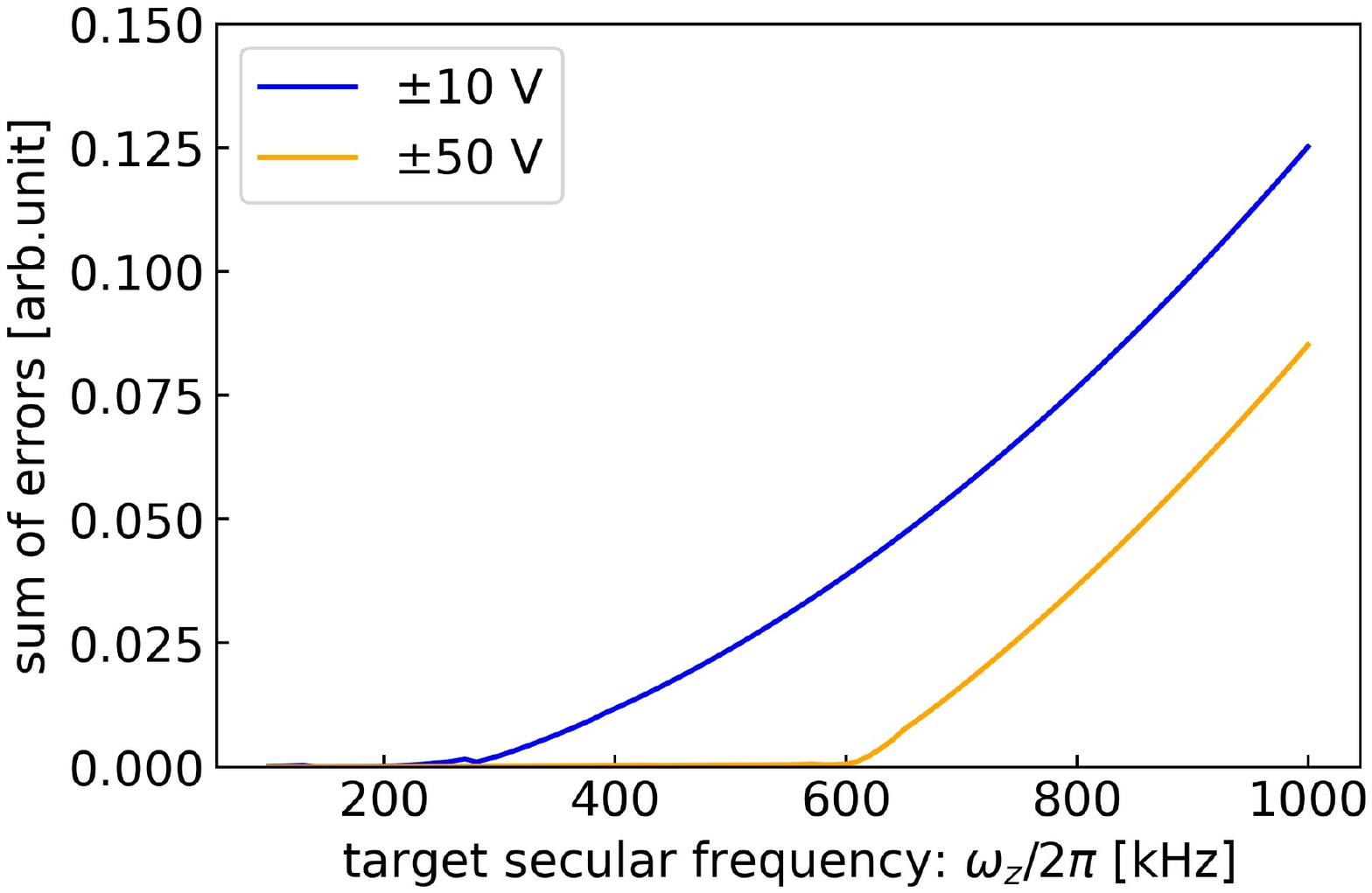}\caption{\label{fig:errorestimate}
Estimation of the optimization error when the output voltage restriction of $\pm10$~V (blue curve) and $\pm50$~V (orange curve). The error is calculated as the difference between the target potential and the potential generated by the optimized voltage sets.}
\end{figure}

The disagreement between the target secular frequency and achieved secular frequency may be attributed to the difference between the objective function values in the optimization process and the potentials calculated from the optimized voltage sets.
This difference arises due to the constraints imposed by the output voltage range.
To evaluate this, we calculated the sum of errors within the optimization range for the target secular frequencies varying from 100~kHz to 1~MHz in increments of 10~kHz.
The results are shown in Fig.~\ref{fig:errorestimate}.
The range where the sum of the errors remains nearly zero is broader for the $\pm50$~V restriction compared to the $\pm10$~V restriction. 
Furthermore, at the target frequency of $2\pi\times$500~kHz, the sum of errors is significantly smaller for the $\pm50$~V restriction than for the $\pm10$~V restriction.
These calculated results correlate well with the experimental results shown in Fig.~\ref{fig:Secular frequency}.

The improvement in secular frequency obtained by varying the range of output voltages in the optimization corresponds to the fact that under ideal conditions, when the voltage applied to all electrodes is scaled by a factor of $s$, the secular frequency increases by a factor of $\sqrt{s}$.
Moreover, for a given electrode geometry, an optimized voltage set remains optimal for any target secular frequency when scaled accordingly.
This removes the need for repeated optimizations, enabling efficient voltage determination through simple scaling and significantly reducing computational costs.

Motional heating of the ions during the shuttling operations is another important issue of  the QCCD architecture.
The scope of the present study is to confirm the increase in the secular frequency by extending  the DAC output range in an actual setup, and experimental evaluation of the noise characteristics will be performed as the next step.

The THD of this system determined by the specification of DAC and OP-amp is approximately -88~dB at 10~kHz.
We estimated that this is at a sufficient level for application to the ion shuttling operations. 
Although accurate evaluation of THD at this level (< -60~dB) by actual measurement is very challenging, we believe that the system works without major problems within the scope of this experiment.

For large-scale QCCD, increasing the number of channels, that is, the number of DACs, is essential.
However, the number of DACs that can be connected to an SoC is constrained by the number of the I/O pins, hindering it to accommodate all the DACs required for QCCD with a single SoC.
Therefore, multiple DAC systems must be integrated to expand the number of channels.
We have explored a system that enables the synchronization of multiple controllers using a time-based event system~\cite{10821043}.
In the future, we aim to control the DAC system presented here within this synchronization framework to facilitate QCCD implementation.

\section{Conclusion}\label{sec:Conclusion}
In this study, we developed an FPGA-based DAC system with an analog output voltage range of $\pm50$~V and successfully demonstrated its application to ion transport operations.
We used quadratic programming to optimize voltage sets for ion-shuttling operations with a constant secular frequency under the restrictions of DAC output ranges.
Our experimental results demonstrate that the DAC system can generate potentials with a secular frequency of $2\pi\times$500~kHz. 
During the shuttling operation, the desired potential at each transport step was validated by measuring the secular frequency of the trapped ions. 
The secular frequency depends on the configuration of trap and the ion species; however, a $\pm50$~V output DAC is effective in increasing the secular frequency, which contributes to ion shuttling operations.

\begin{acknowledgments}
We thank Koichiro Miyanishi for his contributions to planning and coordinating this collaborative research, formulating and specifications of the initial version of the DAC, and preparing the experiments using the initial version of the DAC.
SM, TM, and RO thank Atsushi Noguchi for valuable discussions on the DAC specifications.
This work was supported by JST Moonshot R$\And$D  (Grant Number JPMJMS2063), JST COINEXT (Grant Number JPMJPF2014), JST Moonshot R$\And$D (Grant Number JPMJMS226A), and MEXT Q-LEAP (Grant Number JPMXS0120319794).
\end{acknowledgments}

\section*{data availability statement}
The data that support the findings of this study are available
from the corresponding author upon reasonable request.

\section*{References}
\bibliography{50V_DAC_development_main}
\
\end{document}